\documentclass{aa}  

\usepackage{graphicx}
\usepackage{txfonts}
\usepackage{natbib,twoopt}
\usepackage{amsmath}
\usepackage[nopatch]{microtype}
\usepackage{booktabs}
\usepackage{multirow}
\usepackage{hyphenat}
\usepackage{ulem}
\usepackage{amssymb}
\usepackage{threeparttable}
\usepackage{color,soul}
\usepackage{xcolor}
\usepackage[colorlinks=True,allcolors=blue]{hyperref}

\newcommand{\ergs}{erg~s$^{-1}$}

\newcommand{\msun}{M$_\odot$}
\def\degs{\ifmmode ^{\circ}\else$^{\circ}$\fi}
\def\amin{\ifmmode ^{\prime}\else$^{\prime}$\fi}
\def\asec{\ifmmode ^{\prime\prime}\else$^{\prime\prime}$\fi}
\def\fss{\hbox{$.\!\!^{\rm s}$}}        
\def\farcs{\hbox{$.\!\!^{\prime\prime}$}}  
  
\def\h{$^{\rm h}$}
\def\m{$^{\rm m}$}

\def\rsun{R$_\odot$}
 
\def\psr{J0312}
\def\psrii{J1627}
\def\hip{HiPERCAM}
\def\gtc{Gran Telescopio Canarias}
\def\ps{Pan-STARRS}
\def\sdss{SDSS}
\def\xmm{\textit{XMM-Newton}}
\def\fermi{\textit{Fermi}}
\def\sw{\textit{Swift}}
\def\gaia{\textit{Gaia}}
\def\atnf{\url{https://www.atnf.csiro.au/people/pulsar/psrcat/}}
\newcommand{\flux}{erg~s$^{-1}$~cm$^{-2}$}
\newcommand{\kirr}{erg~cm$^{-2}$~s$^{-1}$~sr$^{-1}$}
\def\ergs{erg~s$^{-1}$}

\begin{document} 

\titlerunning{The black widow pulsars J0312$-$0921 and  J1627+3219} 
\authorrunning{Bobakov et al.}

\title{Studying the black widow pulsars \\ PSR J0312$-$0921 and PSR J1627$+$3219  in the optical and X-rays}

\author{
A. V. Bobakov\inst{1}\thanks{E-mail: bobakov\_alex@mail.ru},
A. Kirichenko\inst{2,1},
S. V. Zharikov\inst{2},
D. A. Zyuzin\inst{1}, \\
A. V. Karpova\inst{1},
Yu. A. Shibanov\inst{1},
T. Begari\inst{3}
}

\institute{
Ioffe Institute, 26 Politekhnicheskaya, St. Petersburg, 194021,  Russia
\and
Instituto de Astronom\'ia, Universidad Nacional Aut\'onoma de M\'exico, Apdo. Postal 877, Ensenada, Baja California, M\'exico, 22800
\and
American Association of Variable Star Observers, 185 Alewife Brook Parkway, Suite 410, Cambridge, MA 02138, USA
}

\date{Received ..., 2025; accepted ..., 2025}

\abstract
{PSR J0312$-$0921 and PSR J1627$+$3219 are black widow pulsars with orbital periods of 2.34 and 3.98 hours.
They were recently detected in the radio and $\gamma$-rays.}
{Our goals were to estimate the fundamental parameters of both binary systems and their components.}
{We performed first phase-resolved multi-band photometry of both objects with the 10.4 m Gran Telescopio Canarias and fitted the obtained light curves with a model assuming direct heating of the companion by the pulsar.
Archival X-ray data obtained with the \sw\ and \xmm\ observatories were also analysed.}
{For the first time, we firmly identified both systems in the optical.
Their optical light curves show a rather symmetric single peak per orbital period and a peak-to-peak amplitude of $\gtrsim$2 mag. 
We also identified the X-ray counterpart to J1627$+$3219, and for J0312$-$0921 we set an upper limit on the X-ray flux.
}
{We estimated the masses of the pulsars, companion temperatures and masses, Roche lobe filling factors, orbital inclinations, and the distances to both systems. 
PSR J0312$-$0921 has a very light companion ($\approx$0.02~\msun) that possibly has one of the lowest night-side temperatures of the known black widow systems ($\approx$1600 K). We find that the distances to J0312$-$0921 and J1627$+$3219 are about 2.5 and 4.6 kpc, respectively. This likely explains their faintness in X-rays. The X-ray spectrum of PSR J1627$+$3219 can be described by a power-law model, and its parameters are compatible with those obtained for other black widows.}


   \maketitle
%

\section{Introduction}
\label{sec:introduction}

Black widows (BWs) form a subclass of the so-called spider systems, which are binary millisecond pulsars (MSPs) with short orbital periods ($P_b \lesssim 1$ d) and low-mass companions ($M_{\rm c} \lesssim 0.05$~\msun\ for BWs; \citealt{roberts2013}).
The companion in these systems is heated and evaporated by the pulsar high-energy radiation and the wind of relativistic particles, which can cause eclipses of the pulsar radio emission.
BW pulsars are recycled via the accretion of matter and angular momentum from main-sequence companion stars during the low-mass X-ray binary stage \citep{Bisnovatyi-Kogan1974,alpar1982}.
In addition, BWs have been proposed as potential progenitors of isolated MSPs \citep[e.g.][]{ginzburg&quataert2020,guo2022}.
However, the formation pathway of such systems is not yet completely clear \citep{chen2013,benvenuto2014,ginzburg&quataert2021,guo2024}.

To date, 49 confirmed BWs have been discovered in the Galactic field, mainly through observations in the radio and $\gamma$-rays \citep{spidercat}. 
These observations can provide information on the source position, pulsar mass function, spin period, binary period, and distance. 
Optical studies of BW systems are crucial for obtaining other fundamental parameters such as the companion's surface temperature distribution, Roche-lobe filling factor, and irradiation efficiency. 
Moreover, they are useful for measuring the masses of binary components, orbital inclinations, and distances, especially in cases where these parameters cannot be constrained well from radio or $\gamma$-ray timing observations alone.
Recent optical studies of a sample of BWs provided valuable constraints on their system parameters \citep{draghis2019,matasanchez2023,Bobakov2024}. 

The BW pulsars PSR J0312$-$0921 and PSR J1627$+$3219 (hereafter \psr\ and \psrii) were recently discovered with the Green Bank Telescope and the Five hundred meter Aperture Spherical radio Telescope, respectively, during searches for unassociated \fermi\ Large Area Telescope (LAT) $\gamma$-ray sources \citep{tabassum2023,xmmprop1627,psr-symp}. 
Pulsations with the pulsars' spin periods were also detected in $\gamma$-rays \citep{smith2023}.
The parameters of the systems are presented in Table~\ref{tab:pars}.
We note that the \psr\ orbital period, 2.34 h, is among the shortest of the known BWs \citep[see e.g.][]{swihart2022}.

Here we report the results of the first multi-band optical observations of \psr\ and \psrii\ with the 10.4 m \gtc\ (GTC).
In addition, we analysed archival X-ray observations carried out with the \sw\ X-Ray Telescope (XRT) and \xmm\ observatory. 
The paper is organised as follows: Optical observations and data reduction are described in Sect.~\ref{sec:data}, and the modelling of the light curves in Sect.~\ref{sec:lc-mod}.
Analysis of the X-ray data is presented in Sect.~\ref{sec:x-rays}.
Discussion and conclusions are given in Sect.~\ref{sec:discussion}.  

\begin{table*}
\renewcommand{\arraystretch}{1.2}
\caption{\psr\ and \psrii\ parameters.}
\label{tab:pars}
\begin{center}
\begin{tabular}{lcc}
\hline
\hline
PSR                                                            & \psr                                 & \psrii \\
\hline
Right ascension $\alpha$ (J2000)                               & 03\h12\m06\fss21465615(2)            & 16\h27\m52\fss9985(5) \\
Declination $\delta$ (J2000)                                   & $-$09\degs21\amin56\farcs55324461(5) & +32\degs18\amin26\farcs643(8) \\
Galactic longitude $l$, deg                                    & 191.510                              & 52.970 \\
Galactic latitude $b$, deg                                     & $-$52.378                            & 43.209 \\
Spin period $P$, ms                                            & 3.7043355330862(3)                   & 2.1828338203418(5) \\
Spin period derivative $\dot{P}$, s s$^{-1}$                   & 1.9723(6)$\times$10$^{-20}$          & 5.478(4)$\times$10$^{-21}$ \\
Orbital period $P_b$, d                                        & 0.0974588765(11)                     & 0.165880827(4) \\
Dispersion measure (DM) pc cm$^{-3}$                            & 20.5                                 & -- \\ 
Distance $D_{\rm YMW}$, kpc                                    & 0.82                                 & 4.47 \\
Characteristic age $\tau_c\equiv P/2\dot{P}$, Gyr              & 3.0                                  & 6.3\\
Observed spin-down luminosity $\dot{E}$, \ergs\                & 1.5$\times$10$^{34}$                 & 2.1$\times$10$^{34}$ \\
Minimum companion mass $M_{\rm c,\ min}$, \msun                & 0.009                                & 0.022 \\
Mass function $f_M$, \msun                                     & 3.9$\times$10$^{-7}$                 & 6$\times$10$^{-6}$ \\ 
P.m. in RA direction $\mu_\alpha$cos$\delta$, mas~yr$^{-1}$    & 31.7$\pm$1.2                         & 1.6$\pm$1.4 \\
P.m. in Dec direction $\mu_\delta$, mas~yr$^{-1}$              & 8.9$\pm$3.0                          & $-$2.8$\pm$1.5 \\
\hline
\end{tabular}
\end{center}
\smallskip

\small{\textbf{Notes.} Parameters are taken from the web page of the Third \fermi\ LAT catalogue of $\gamma$-ray pulsars (\url{https://fermi.gsfc.nasa.gov/ssc/data/access/lat/3rd_PSR_catalog/3PC_HTML/}) and the Australia Telescope National Facility Pulsar catalogue (\citealt{atnf}; \atnf).
Numbers in parentheses denote 1$\sigma$ uncertainties relating to the last significant digit quoted.
P.m. $\equiv$ proper motion.
$D_{\rm YMW}$ is the DM distance calculated using the YMW16 \citep{ymw2016} model for the distribution of free electrons in the Galaxy. For \psrii, the value is taken from \citet{smith2023}.
The spin-down luminosity is calculated assuming the canonical moment of inertia of 10$^{45}$ g~cm$^2$.
The minimum companion mass is calculated assuming a system inclination of $i=90$\degs\ and a canonical pulsar mass of $M_{\rm p}=1.35$~\msun\ \citep[similarly to e.g.][]{spidercat}.
The mass function for \psr\ is calculated from the orbital parameters.}
\end{table*}

\begin{figure}[th]
\begin{minipage}[h]{0.93\linewidth}
\centering{\includegraphics[width=1.0\textwidth, trim={0 0 0 0.7cm}, clip]{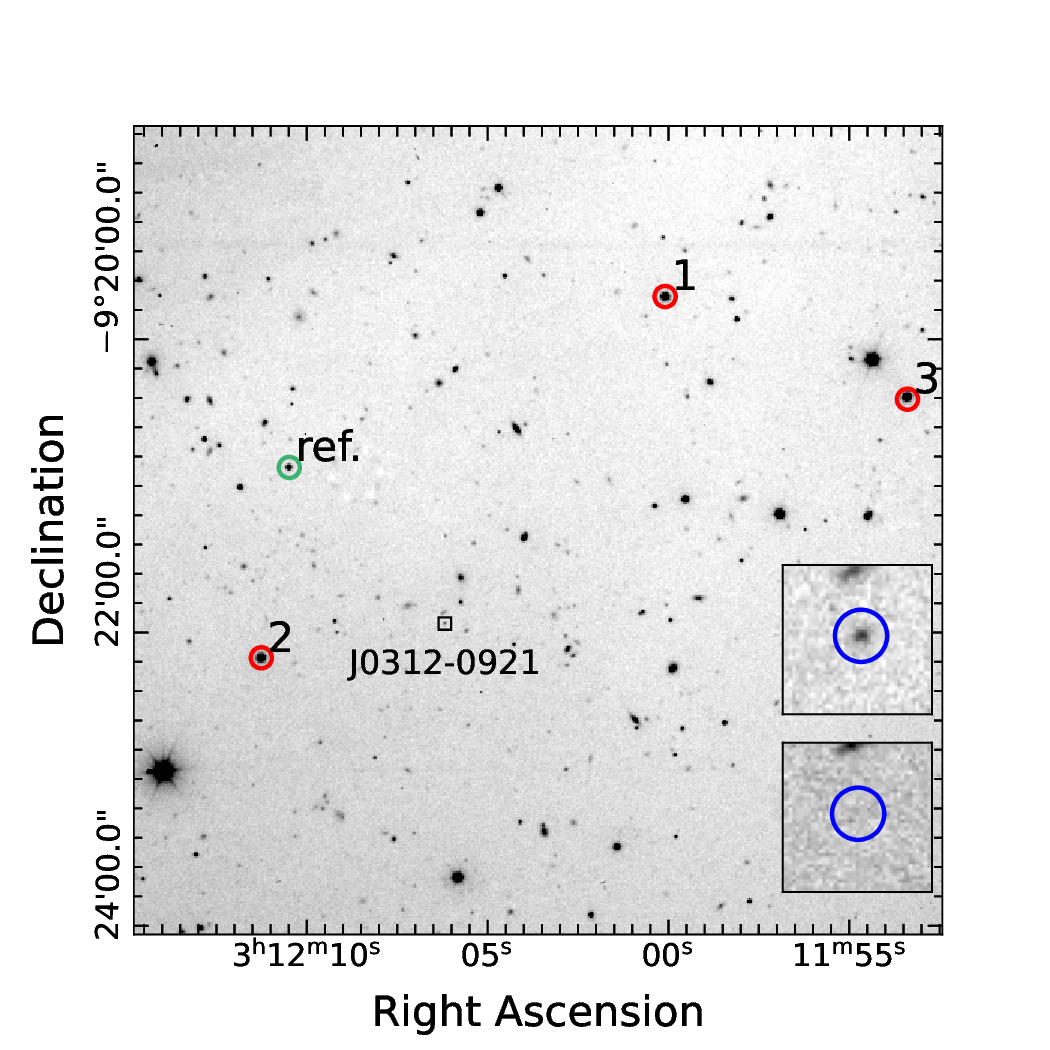}}
\end{minipage}

\begin{minipage}[h]{0.93\linewidth}
\centering{\includegraphics[width=1.0\textwidth, trim={0 0.5cm 0 2.3cm}, clip]{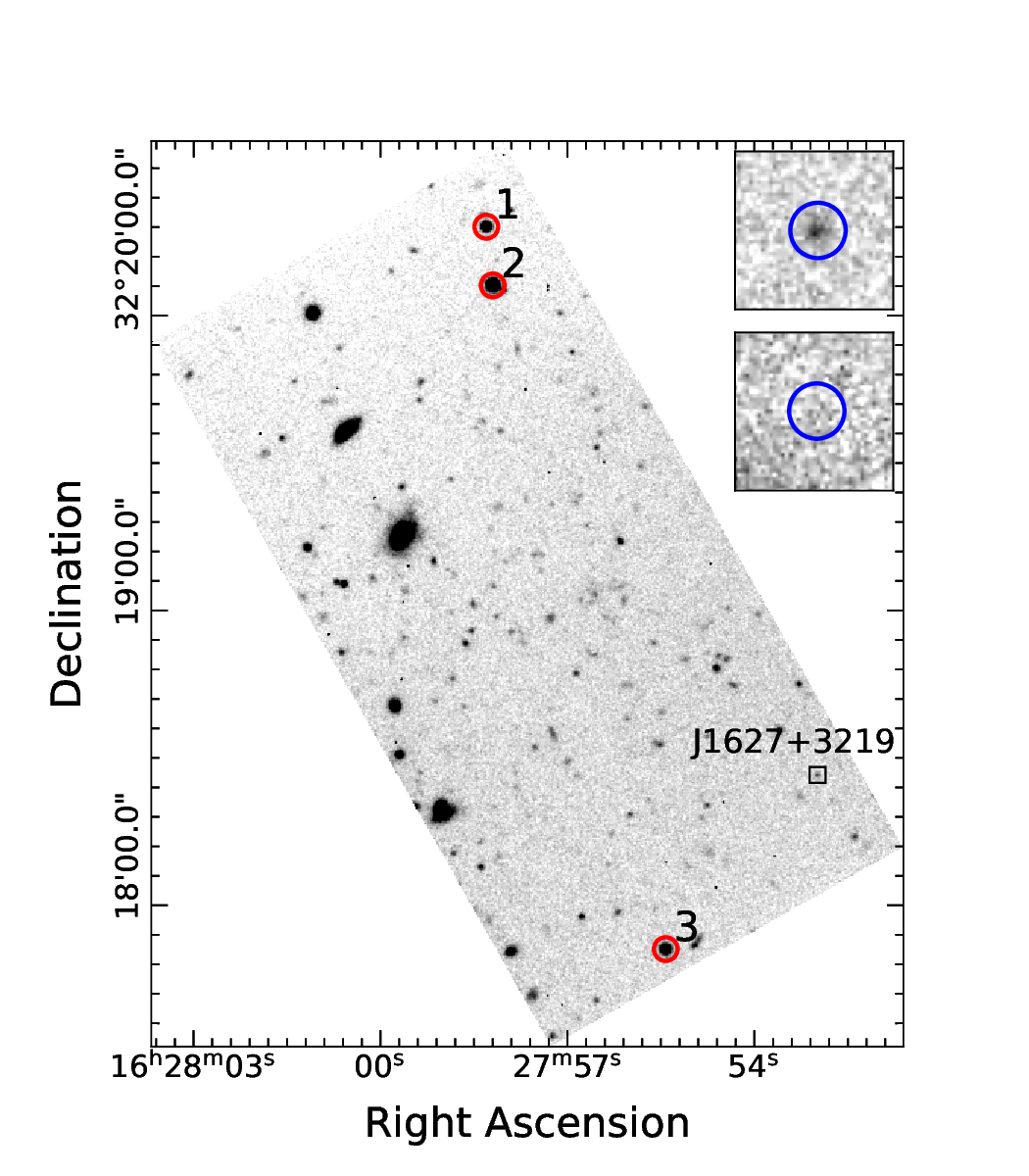}}
\end{minipage}

\caption{Optical images of the fields of the two pulsars. 
\textit{Top panel}: 3\farcm05~$\times$~3\farcm05 image of the \psr\ field obtained with the GTC/OSIRIS in the $r'$ band. 
\textit{Bottom panel}: 2\farcm8~$\times$~1\farcm4 FoV of the GTC/HiPERCAM  in the $g_s$ band containing \psrii.
The pulsars' vicinities are shown with the boxes and enlarged in the insets to demonstrate the maximum (top) and minimum (bottom) brightness phases of their companions, whose positions are marked by blue circles.
The positions of the stars used for the photometric calibration are marked by red circles.
The star used to account for the changing weather conditions is shown with the green circle in the top panel. For J1627, for this purpose we used the same stars as for the photometric calibration.}
         \label{fig:field}
\end{figure}

\section{Optical observations and data reduction}
\label{sec:data}

\begin{table}
\renewcommand{\arraystretch}{1.2}
\begin{center}
\caption{Log of observations.}
\label{tab:log}
\begin{tabular}{ccccc}
\hline
\hline
Date      & Filter & Exposure            & Airmass   & Seeing, \\  
yy/mm/dd  &        & time, s             &           & arcsec  \\ 
\hline
\multicolumn{5}{c}{\textbf{\psr}} \\
2023/10/10 & $g'$  & 200 $\times$ 18     & 1.3--1.4  & 0.6$-$0.9  \\
           & $r'$  & 180 $\times$ 20     &           &  \\
2023/10/11 & $r'$  & 180 $\times$ 18     & 1.3--1.6  & 0.7--0.8 \\
           & $i'$  & 120 $\times$ 18     &           &  \\
\hline
\multicolumn{5}{c}{\textbf{\psrii}} \\
2024/07/29 & $u_s$ & 121.6 $\times$ 126 & 1.0--2.0   & 0.7--2.5 \\
           & $g_s$ & 60.8 $\times$ 252  &            &           \\
           & $r_s$ & 60.8 $\times$ 252  &            &           \\
           & $i_s$ & 60.8 $\times$ 252  &            &           \\
           & $z_s$ & 121.6 $\times$ 126 &            &           \\       
\hline
\end{tabular}
\end{center}
\end{table}

\subsection{\psr}
\label{subsec:0312}

The phase-resolved photometric observations\footnote{Proposal GTC14-23BMEX, PI A. Kirichenko} of the \psr\ field were performed during two observing runs in October 2023 in the Sloan $g'$, $r'$, and $i'$ bands with the Optical System for Imaging and low Resolution Integrated Spectroscopy (OSIRIS+) instrument\footnote{\url{https://www.gtc.iac.es/instruments/osiris+/osiris+.php}}. 
The OSIRIS+ field of view (FoV) is 7\farcm8~$\times$~7\farcm8 with a pixel scale of 0\farcs254 in the standard 2$\times$2 binning mode. 

The observations were carried out under photometric weather conditions. To avoid effects from CCD defects, we used 5\asec\ dithering between the individual exposures. Each observing run  (with durations of about 2.7 h and 2.3 h)  covered approximately one orbital period of the system. To optimise the efficiency of observations, in the first run we used the alternating $g'$ and $r'$ bands, while in the second run the BW period was covered in the $r'$ and $i'$ bands. 
The log of observations is presented in Table \ref{tab:log} and the $r'$-band image of the pulsar field is shown in the top panel of Fig.~\ref{fig:field}, where the variability of the pulsar companion is demonstrated in the insets. 

Standard data reduction, including bias subtraction and flat-fielding, was performed using the Image Reduction and Analysis Facility (IRAF) package. 
We applied the L.A.Cosmic algorithm \citep{vandokkum} to remove cosmic-ray events.
For astrometric referencing, we used a single 180 s $r'$-band image and a set of stars from the \gaia\ Data Release 3 catalogue \citep{2023A&A...674A...1G}.
The formal rms uncertainties of the fit were $\Delta \alpha$~$\la$~0\farcs13 and $\Delta \delta$~$\la$~0\farcs14.

For the photometric calibration, we used the Sloan photometric standard SA 112-805  \citep{smith} observed during the same nights as the target. 
Accounting for the extinction coefficients $k_{g'}$ = 0.15(2), $k_{r'}$ = 0.07(1), and $k_{i'}$ = 0.04(1) \citep{extinc}, 
we calculated the zero points $Z_{g'}=28.07(2)$, $Z_{r'} = 28.09(1)$ (2023 October 10), $Z_{r'} = 28.02(1)$ (2023 October 11), and $Z_{i'}=27.58(1)$. We verified these zero points using a set of stars from the Panoramic Survey Telescope
and Rapid Response System (\ps) Data Release 2 (DR2) catalogue (\citealt{flewelling2020}, see Fig.~\ref{fig:field}) and short exposures obtained in each band for the calibration purposes. 
The 3$\sigma$ detection limits during the observations were $g' \approx 26.5$ mag, $r' \approx 25.75$--$25.9$ mag, and $i' \approx 24.6$--$24.8$ mag.
We note that all magnitudes presented in this paper are in the AB system. 

We used the optimal extraction algorithm \citep{optimal_slgorithm}, to extract the target light curves and applied a differential technique to eliminate the variations due to the changing weather conditions using a non-variable bright field star from the \ps\ DR2 catalogue as a reference (Fig.~\ref{fig:field}, top). 
Its magnitude scatters were $\lesssim 0.05$ mag during the runs.

\subsection{\psrii}
\label{subsec:1627}

The phase-resolved multi-band observations of the \psrii\ field were performed on  2024 July  29\footnote{Proposal GTC6-24AMEX, PI A. Kirichenko} with the HiPERCAM instrument\footnote{\url{http://www.gtc.iac.es/instruments/hipercam/hipercam.php}} \citep{dhillon2016,dhillon2018,dhillon2021} simultaneously in five ($u_s$, $g_s$, $r_s$, $i_s$, and $z_s$) high-throughput `Super' SDSS  filters.
The HiPERCAM FoV is 2\farcm8~$\times$~1\farcm4, with the 0\farcs16 pixel scale in the $2\times2$ binning mode. 
The observations were performed under photometric weather conditions. 
The log of observations is presented in Table~\ref{tab:log}. 
The total duration was 4.4 h covering one orbital period.

We performed standard data reduction including bias subtraction, flat-field correction and $z_s$-band fringe removal following the pipeline manual\footnote{\url{https://cygnus.astro.warwick.ac.uk/phsaap/hipercam/docs/html/}}. 
An example of the $g_s$-band individual image is presented in Fig.~\ref{fig:field}, bottom. 
The pulsar companion variability is demonstrated in the two insets.

Using the optimal extraction algorithm \citep{optimal_slgorithm}, we measured instrumental magnitudes of the counterpart and three stars in the FoV whose magnitudes are available in the \sdss\ DR18 catalogue \citep{sdss-18}.
To avoid centroiding problems of the companion during its faint brightness stages, its position was fixed relative to the nearest star when the source was in its maximum brightness phase.
We performed photometric calibration using the GTC atmospheric extinction coefficients $k_{u_s}=0.48$, $k_{g_s}=0.17$, $k_{r_s}=0.1$, $k_{i_s}=0.05$, and $k_{z_s}=0.05$ \citep{dhillon2021} and the spectrophotometric standards Feige110 and WD1606+422 \citep{Oke1990-std-ob2, WD1606+442} observed during the same night as the target. 
To eliminate possible systematic errors and to account for photometric zero-point variations during the observation, we used the technique described in \citet{LCcorrTechHonneycutt} and stable \ps\ stars (Fig. \ref{fig:field}) whose magnitude scatters were $\lesssim 0.05$ mag  during the observing run. 
These magnitudes were compared with their catalogue values. This allowed us to verify the calibration, which resulted in the final zero-points $Z_{u_{s}}=27.62\pm0.05$, $Z_{g_{s}}=28.87\pm0.01$, $Z_{r_{s}}=28.46\pm0.02$, $Z_{i_{s}}=28.00\pm0.02$, $Z_{z_{s}}=27.67\pm0.02$.  
Due to the weather conditions, detection limits varied in the ranges of $u_s \approx$  24.6--25.6 mag, $g_s \approx$ 24.9--25.7 mag, $r_s \approx$ 24.3--25.2 mag, $i_s \approx$ 23.9--24.7 mag, and $z_s \approx$ 23.6--24.4 mag. 

The light curves of \psrii\ were obtained in the $g_s$, $r_s$, $i_s$, and $z_s$ bands while in the $u_s$ band its brightness was below the \hip\ sensitivity limit.

\section{Optical light curves and the system parameters}
\label{sec:lc-mod}

\begin{figure*}
\begin{center}
\includegraphics[width=0.49\textwidth, clip]{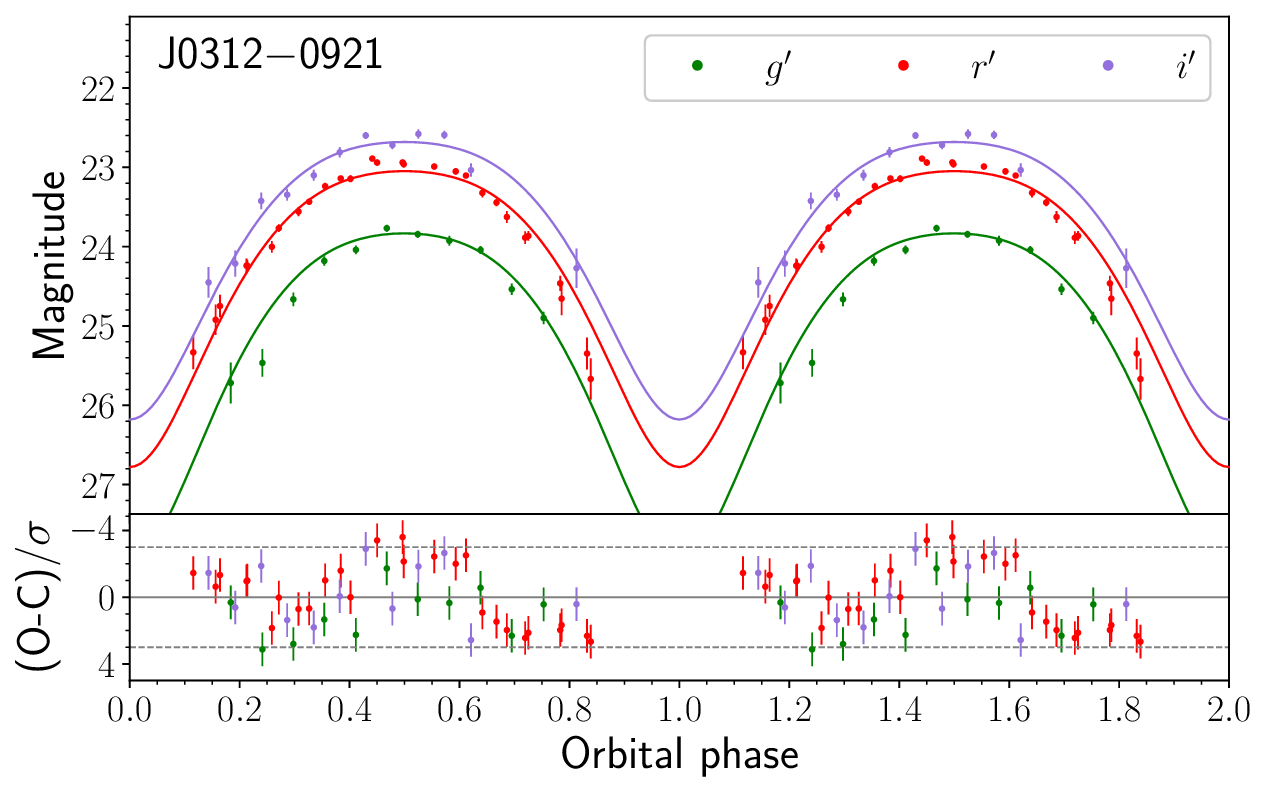}
\includegraphics[width=0.49\textwidth, clip]{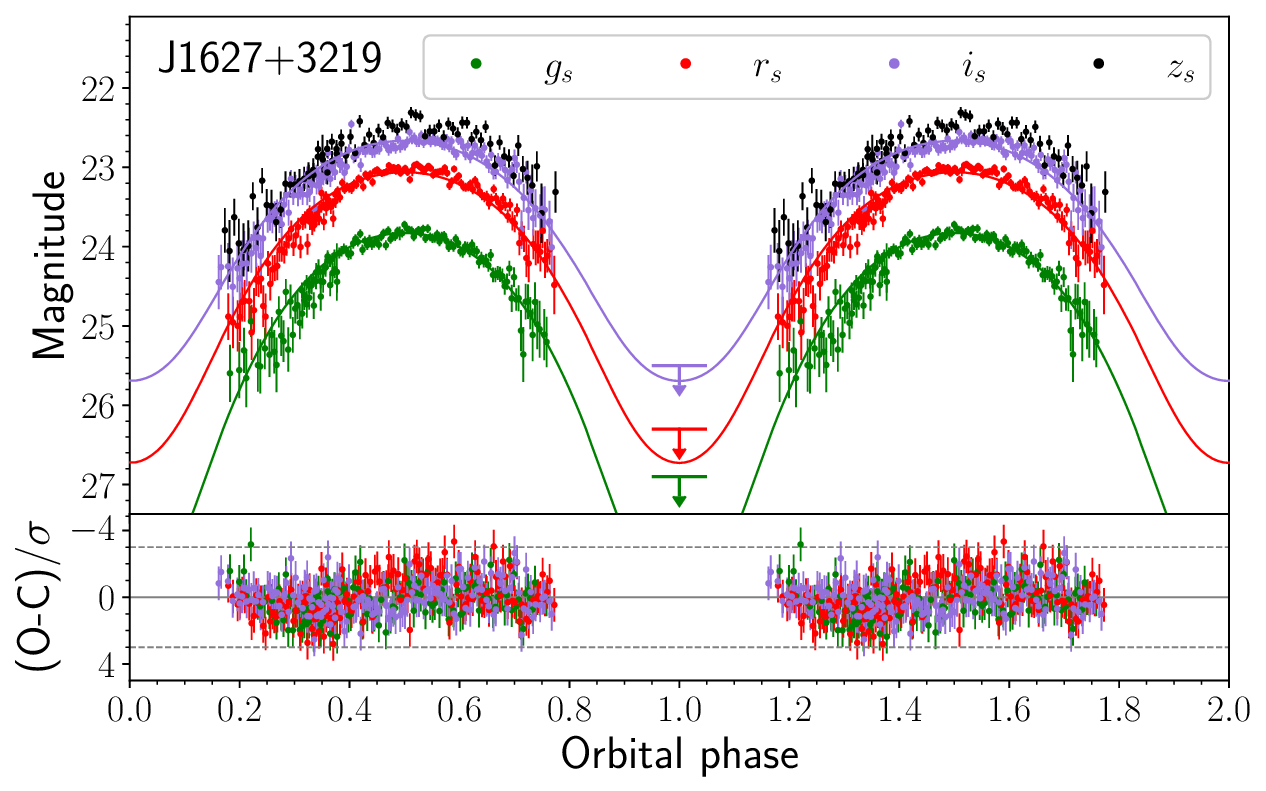}
\end{center}
\caption{Light curves of \psr\ (left) and \psrii\ (right) folded with the orbital periods and the best-fitting models (solid lines). 
Two periods are shown for clarity. 
The orbital phases $\phi=0.0$ correspond to the minima of the models' brightness. 
Panels show residuals calculated as the difference between the observed ($O$) and calculated ($C$) magnitudes for each data point in terms of the magnitude error, $\sigma$.
Dashed lines correspond to 3$\sigma$ levels.}  
\label{fig:LCs}
\end{figure*}

\begin{table}
\renewcommand{\arraystretch}{1.2}
\caption{Light-curve fitting results.}
\label{tab:fit} 
\begin{center}
\begin{tabular}{lcc}
\hline
\hline
Fitted parameters                              & \psr\     & \psrii\    \\
\hline
Pulsar mass $M_{\rm p}$, \msun                 &   2.6$_{-0.9}^{+0.4}$    &  2.7$_{-1.0}^{+0.3}$ \\
Mass ratio $q$ =  $M_{\rm c}/M_{\rm p}$        &   0.008(1)   &  0.017(2)\\             
Distance $D$, kpc                              &   2.5(2)     &  4.6(2)\\
Reddening $E(B-V)$, mag                        &   0.18(4)    &  0.051(25)\\
`Night-side' temperature $T_{\rm n}$, K        &   1600$_{-600}^{+200}$    &  2730(300) \\
Inclination $i$, deg                           &   43$_{-7}^{+13}$      &  54$_{-4}^{+10}$ \\
Roche lobe filling factor $f_x$                &   0.94$_{-0.10}^{+0.05}$    &  0.71(7) \\
Irradiation factor $K_{\rm irr}$,              &   0.8(2)     &  2.4(2)  \\
$10^{20}$ \kirr                                &           & \\
$\chi^2$/d.o.f.                                &   205/44     &  570/448 \\
\hline
Derived parameters                             &           &  \\
\hline
Companion mass $M_{\rm c}$, \msun              &   0.02(1)    &  0.04(2) \\
Companion radius $R_{\rm c,x}$, \rsun          &   0.14(2)    &  0.21(2) \\
Companion radius $R_{\rm c,y}$, \rsun          &   0.11(2)    &  0.15(2)\\
Max `day-side' temp. $T_{\rm d}^{\rm max}$, K  &   5940(200)    &  5570(80)\\
Irradiation efficiency $\eta$                  &   0.31(8)     &  0.64(5) \\
\hline
\end{tabular}
\end{center}

\smallskip
\small{\textbf{Notes.} Irradiation efficiencies $\eta$ are calculated using the observed spin-down luminosities from Table~\ref{tab:pars}. $R_{\rm c,x}$ and $R_{\rm c,y}$ are the radii of the ellipsoidal companion. The former is along the line passing through the centres of the binary components. $K_{\rm irr}=\frac{\eta\dot{E}}{4\pi^2 R_{\rm p}^2}$, where $R_{\rm p}=12$ km is the pulsar radius. D.o.f. $\equiv$ degrees of freedom.
The errors of the derived parameters were calculated as $
\Delta f = \sqrt{\sum_{i=1}^n \left( 
\frac{\partial f}{\partial x_i}\bigg|_{x_i = x_i^{\rm best}} 
\, \Delta x_i \right)^2},
$ where $f$ is the function corresponding to the derived parameter, $n$ is the number of fitted parameters, and $\Delta {x_i}$ is the absolute value of the maximum error of the parameter $x_i$ at its best value.}
\end{table}

The resulting barycentre-corrected light curves of \psr\ and \psrii\ folded with their orbital periods (Table~\ref{tab:pars}) are presented in Fig.~\ref{fig:LCs}. 
To estimate the system parameters, we fitted the light curves using the direct heating model consisting of a neutron star as the primary irradiating a low-mass companion as the secondary.
The emission of each surface element of the secondary is approximated by a black-body spectrum with an effective temperature varying from element to element. 
We note that for \psrii\ we used only $g_s$-, $r_s$-, and $i_s$-band points since the model cannot describe data in the wider spectral range from the $g_s$ to $z_s$ band using the black body approximation for an element irradiation.
Details of the model can be found in \citet{zharikov2013,zharikov2019} and \citet{Kirichenko2024MNRAS}. The model takes into account the irradiation  of the secondary $ T_{\rm d} = (T_{\rm n}^4 + \sigma^{-1} \text{cos}(\alpha_n)\Omega K_{\rm irr})^{1/4}$ as it was described in \citet{zharikov2019}\footnote{$\alpha_n$ is the angle between the incoming flux and the normal
to the secondary surface, $\Omega =\pi R_{\rm p}^2/a^2$ is the solid angle from which the
pulsar is visible from the secondary, and $a$ is the orbit separation, $K_{\rm irr}$ is the effective irradiation factor of the secondary, and $\sigma$ is the Stefan-Boltzmann constant.}, the quadratic law of the limb-darkening with coefficients from \citet[Eq. 2 therein]{2012A&A...546A..14C, 2013A&A...552A..16C}, and the gravity darkening from \citet[Eq. 5.4 therein]{2018maeb.book.....P}.

The fitted parameters are the interstellar reddening ($E(B-V)$), the distance ($D$), the pulsar mass ($M_{\rm p}$), the component mass ratio ($q$), the orbit inclination ($i$), the effective irradiation factor ($K_{\rm irr}$), which defines the heating of the companion, the companion Roche lobe filling factor ($f$), defined as a ratio of distances from the centre of mass of the secondary to the star surface and to the Lagrange point $L_{1}$, and the companion `night-side' temperature ($T_{\rm n}$). 
We used the mass functions from the radio observations (Table~\ref{tab:pars}) to link the companion mass $M_{\rm c}$ (or the mass ratio $q=M_{\rm c}/M_{\rm p}$), $M_{\rm p}$, and $i$.
The minimum of the $\chi^2$ function was determined using the gradient descent method. This approach was preferred as it considerably reduces the computational effort required for the minimisation. The parameter uncertainties were calculated following the method
proposed by \citet{1976ApJ...208..177L}.
The results are presented in Table \ref{tab:fit} and the best-fitting models for both objects are shown by solid lines in Fig.~\ref{fig:LCs}. 

To check whether the model predictions near the \psrii\ minimum brightness phase are adequate, we combined 20 images at $\phi \approx 0.00 \pm 0.05$. 
Nevertheless, the target was not detected in any band.
The corresponding detection limits were $g_s \approx 26.9$ mag, $r_s \approx 26.3$ mag, and $i_s \approx 25.5$ mag, which is compatible with the model. 
For \psr, the number of data points near the minimum brightness phase is not enough to significantly improve the detection limits derived in Sect.~\ref{subsec:0312}.

\section{X-ray data}
\label{sec:x-rays}

\begin{figure}
    \centering
    \includegraphics[width=0.93\linewidth]{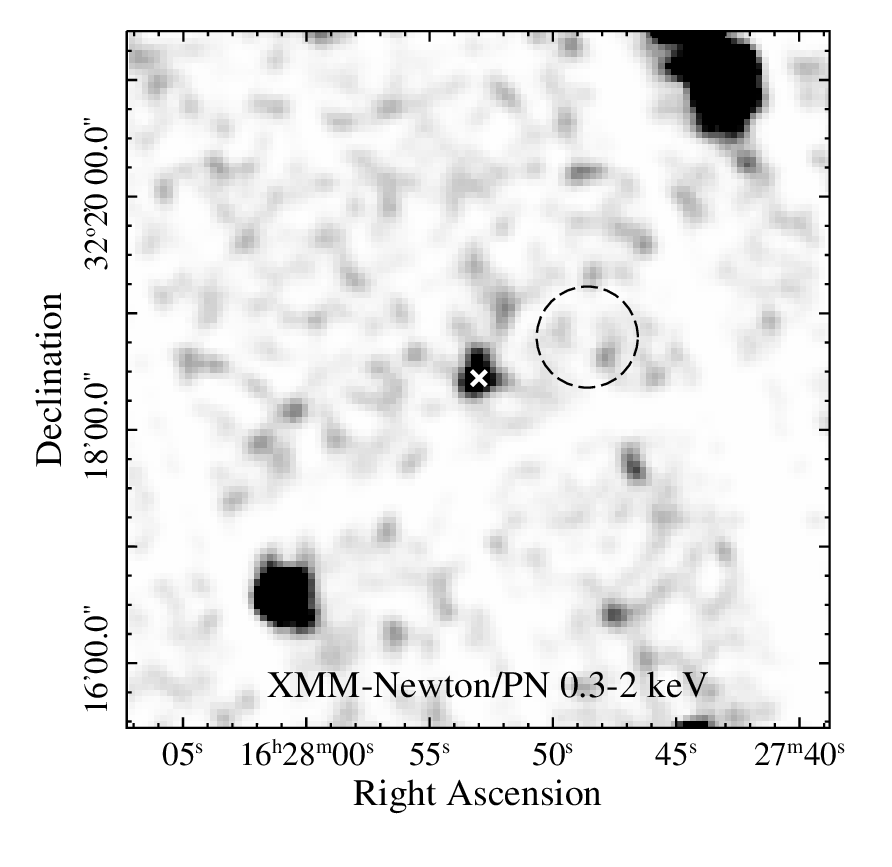}
    \caption{6\amin~$\times$~6\amin\ \xmm/PN image of the \psrii\ field in the 0.3--2 keV band.  
    The `X' symbol marks the pulsar timing position from Table~\ref{tab:pars}.
    The dashed circle shows the region chosen for the background extraction.
    }
    \label{fig:xmm}
\end{figure}

The \psr\ and \psrii\ fields were observed with \sw/XRT several times between 2010 and 2019 with total exposure times of 5.4 and 8.1 ks, respectively. 
No statistically significant sources were detected at the pulsars' positions.
Using the Living \sw-XRT Point Source (LSXPS) Upper limit server\footnote{\url{https://www.swift.ac.uk/LSXPS/ulserv.php}} \citep{evans2023}, we derived the 3$\sigma$ upper limits on the sources' count rates of CR$^{\rm \psr} = 2.2\times10^{-3}$~cts s$^{-1}$ and CR$^{\rm \psrii} = 1.2\times10^{-3}$ cts s$^{-1}$.
This corresponds to the unabsorbed fluxes in the 0.5--10 keV range $F_X^{\rm \psr}\approx7.9\times10^{-14}$ \flux\ and $F_X^{\rm \psrii}\approx3\times10^{-14}$ \flux.
Here we assume a power law model with the photon index $\Gamma = 2.5$, which is the average value for the BW family \citep{swihart2022}.
The absorbing column densities were $N^{\rm \psr}_{\rm H}=1.6 \times 10^{21}$~cm$^{-2}$ and $N^{\rm \psrii}_{\rm H}=4.5 \times 10^{20}$~cm$^{-2}$.
They were calculated using the reddening values obtained for \psr\ and \psrii\ from the optical fits (Table~\ref{tab:fit}) and the empirical relation from \citet{foight2016}. 

The \psrii\ field was also observed\footnote{ObsID 0902730101, PI P. Saz Parkinson} with \xmm\ on 2023 February 10 with duration of 38.5 ks.
The European Photon Imaging Camera-Metal Oxide Semiconductor (EPIC-MOS) and EPIC-pn (PN hereafter) detectors were operated in the full frame mode with the thin filter. 
The \textit{XMM-Newton} Science Analysis Software (XMM-SAS) v.22.1.0 was utilised for data reduction. 
The data were reprocessed using the \texttt{emproc} and \texttt{epproc} routines.
Unfortunately, the high-energy light curves extracted from the FoVs of detectors showed a significant number of strong background flares.
We filtered them out applying the \texttt{espfilt} task.
This resulted in effective exposure times of 18.4, 18.8, and 11.2 ks for the MOS1, MOS2, and PN detectors, respectively.
Using the \texttt{edetect\_chain} tool\footnote{\url{https://www.cosmos.esa.int/web/xmm-newton/sas-thread-src-find}} and data from all detectors, we performed source detection.
A weak X-ray source, the likely counterpart of \psrii, was detected at the pulsar timing position 
(Fig.~\ref{fig:xmm}).
Its coordinates are $\alpha_X$~=~16\h27\m53\fss15 and $\delta_X$~=~32\degs18\amin26\farcs0, and its position uncertainty is 2\asec\ (which combines the statistical uncertainty of 1\farcs6 and the absolute astrometry accuracy for \xmm/EPIC\footnote{\url{https://xmmweb.esac.esa.int/docs/documents/CAL-TN-0018.pdf}.} of 1\farcs2). 
We also found that the source is not detected in the MOS1 or MOS2 data alone likely due to lower efficiency of the MOS detectors in the soft energy band ($\lesssim 2$ keV) in comparison with the PN camera.

We extracted the source spectrum from the PN data using 10\asec-radius aperture around its position using the \texttt{evselect} tool.
For the background, the 26\asec-radius circle was chosen (Fig.~\ref{fig:xmm}).
The redistribution matrix and ancillary response files were created by the \texttt{rmfgen} and \texttt{arfgen} tasks.
We obtained 30.6 net counts in the 0.2--10 keV band and grouped the spectrum to ensure at least 1 count per energy bin.
It was fitted in the X-Ray Spectral Fitting Package (XSPEC) v.12.15.0 \citep{xspec}, applying the $W$-statistics appropriate for Poisson data with Poisson background\footnote{\url{https://heasarc.gsfc.nasa.gov/xanadu/xspec/manual/XSappendixStatistics.html}} and the power law model.
The \texttt{tbabs} model with the \texttt{wilm} abundances \citep{wilms2000} was also included to account for the interstellar absorption.
The absorbing column density was fixed at the value mentioned above.
As a result, we derived the photon index $\Gamma=3.3\pm0.5$, the unabsorbed flux in the 0.5--10 keV band $F_X^{\rm \psrii} = 4.2^{+1.6}_{-1.2} \times 10^{-15}$ \flux\ and $W/{\rm d.o.f.}=25/24$ 
(uncertainties are at 1$\sigma$ confidence). 
The flux value is in agreement with the upper limit obtained from the \sw\ data.

\section{Discussion and conclusions}
\label{sec:discussion}

We carried out the first multi-band time-series optical photometry of \psr\ and \psrii. 
The light curves of both systems are rather symmetric and have one peak per orbital period with a peak-to-peak amplitude of $\gtrsim 2$ mag, which is typical for the BW family \citep{swihart2022,matasanchez2023}.
The curves can be described by the direct heating model. We note that there is a hint of some deviation from the model seen in the residuals for both objects. This can be caused by different effects, such as cold spots on the donor star surface \citep{clark2021}, heating by the intra-binary shock \citep{romani&sanchez2016}, and heat redistribution over the companion surface via convection and diffusion \citep{kandel&romani2022, voisin2020}. 
However, the significance of these possible features is low, and further investigations are necessary to confirm or reject them.

The companions in both systems have very low masses ($M_{\rm c}^{\rm \psr} =0.02$~\msun\ and $M_{\rm c}^{\rm \psrii} =0.04$~\msun), which is typical for the BW population \citep{swihart2022}.
Their temperatures are also compatible with those obtained for BWs \citep{matasanchez2023}; the \psr\ companion may have one of the lowest `night-side' temperatures  ($\approx$1600 K) of the known BWs.

For a significant number of spider pulsars, dispersion measure (DM) distances are compatible with the parallax distances or are lower than them (see Fig. 5 in \citealt{koljonen&linares2023}). The same situation occurs for our targets: for \psrii\ the distance provided by the optical fit (4.6 kpc) is close to the DM one (4.5 kpc), while for \psr\ it is much greater (2.5 vs 0.8 kpc).

The reddening value for \psrii\ obtained from the optical fit (Table~\ref{tab:fit}) is in agreement with the maximal $E(B-V) \approx 0.04$ mag derived from the 3D dust map of \citet{dustmap2019}.
For \psr, the maximal $E(B-V)$ is about 0.1 mag, which is less than the best-fitting value.
However, inspection of the map shows that the interstellar medium in the \psr\ circumstance is rather non-uniform, providing a $E(B-V)$ of up to $\approx 0.2$ mag. 
There are no main-sequence stars at distances $\gtrsim 2.1$ kpc, and the star density in the pulsar field is low.
Thus, the map results can be dubious.
In addition, the uncertainties of the model reddening are quite large.

The \psr\ and \psrii\ proper motions derived from the radio timing are $\mu^{\rm \psr} = 33.0\pm 1.4$ mas~yr$^{-1}$ and $\mu^{\rm \psrii} = 3.2 \pm 1.4$ mas~yr$^{-1}$.
The \psrii\ transverse velocity corresponding to the distance 4.6 kpc from the optical fit is $v_t \approx 70$ km~s$^{-1}$. This value is compatible with typical binary pulsar velocities, 
which are mostly lower than 150 km s$^{-1}$  \citep{2005MNRAS.360..974H}. 
In contrast, \psr\ has 
a transverse velocity of $\approx 400$ km~s$^{-1}$ at 2.5 kpc, which is 
considerably higher than, for example, the $v_t = 326$ km~s$^{-1}$ of the high-velocity MSP PSR B1257+12 \citep{2013MNRAS.433..162Y}.
Moreover, correction for the Shklovskii effect \citep{shklovskii} and acceleration due to differential Galactic rotation \citep{nice&taylor,lynch2018} leads to a negative intrinsic $\dot{P}$ value.
This implies that either the distance or the proper motion is overestimated. 
Formally, the DM distance of 0.8 kpc provides an acceptable positive $\dot{P}$ and $v_t$. 
However, a smaller distance implies a smaller intrinsic flux of the companion, leading to an unreasonably small companion radius of $\sim$(25--30)$\times$10$^3$ km.
Reliable measurements of the proper motion require a close distance and/or long time base of observations.
Dispersion measure variations and timing noise can also lead to inaccurate calculations.
For this reason, the \psr\ proper motion may be overestimated similar to that of, for example, BW PSR J1641$+$8049. An updated timing solution for this pulsar yielded a significantly lower proper motion than the previous estimate, allowing us to rule out the spin-up scenario \citep{Kirichenko2024MNRAS}.

According to the \fermi\ LAT 14-Year Point Source Catalog DR~4, the \psr\ and \psrii\ fluxes in the 0.1--100 GeV range are $F^{\rm \psr}_{\gamma} = 5.6(4)\times10^{-12}$~\flux\ and $F^{\rm \psrii}_{\gamma} = 4.0(3)\times10^{-12}$~\flux\ \citep{4fgl-dr4}. 
Using the distances from Table~\ref{tab:fit}, we calculated the corresponding luminosities: $L_\gamma^{\rm \psr} = 4.2\times10^{33}$~\ergs\ and $L_\gamma^{\rm \psrii} \approx 10^{34}$~\ergs. 
The 3$\sigma$ upper limit on the \psr\ X-ray luminosity is $L_X^{\rm J0312} < 5.9\times10^{31}$~\ergs.
The X-ray luminosity of \psrii\ is $L_X^{\rm J1627} \approx 1.1 \times 10^{31}$~\ergs. 
These values are consistent with those of other BWs \citep{swihart2022, koljonen&linares2023}. 
The \psrii\ photon index of $\approx$3.3 is also typical for a BW.
It is high enough to suggest the presence of a thermal component originating from the heated polar caps, although due to the low count statistics we cannot draw any definite conclusions.

The $\gamma$-ray and X-ray efficiencies calculated using the observed spin-down luminosities for \psr\ are $\eta_\gamma^{\rm \psr} = L_\gamma/\dot{E} = 0.28$ and $\eta_X^{\rm \psrii} = L_X/\dot{E} < 4 \times 10^{-3}$ and  for \psrii\ -- $\eta_\gamma^{\rm \psrii} = 0.48$ and $\eta_X^{\rm \psrii} = 5 \times 10^{-4}$.
While the X-ray efficiencies are reasonable for BWs, the $\gamma$-ray and irradiation efficiencies are very high.
This might be explained by the fact that here we used the observed values of the spin-down luminosity, whereas the intrinsic values can be significantly different.
The correction for the acceleration due to differential Galactic rotation is small and does not critically change $\dot{E}$.
The Shklovskii effect for \psrii\ is also negligible.
However, if the pulsars have masses higher than the canonical value of 1.4~\msun, then their true $\dot{E}$ can essentially be higher.
For example, for a pulsar radius of 12--13 km and a mass of 1.7~\msun, which is a lower bound for both objects (see Table~\ref{tab:fit}), the intrinsic $\dot{E}$ will be $\sim$2 times higher\footnote{Here we applied the formula from \citet{Ravenhall&Pethick1994}.}.

The next generation of instruments should help us better constrain the parameters of the systems.

\begin{acknowledgements}

We thank the anonymous referee for useful comments.
The work is based on observations made with the Gran Telescopio Canarias (GTC), installed at the Spanish Observatorio del Roque de los Muchachos of the Instituto de Astrof\'isica de Canarias, on the island of La Palma and on observations obtained with \xmm, a ESA science mission with instruments and contributions directly funded by ESA Member States and NASA. 
This work made use of data supplied by the UK Swift Science Data Centre at the University of Leicester.
This work has made use of data from the European Space Agency (ESA) mission {\it Gaia} (\url{https://www.cosmos.esa.int/gaia}), processed by the {\it Gaia} Data Processing and Analysis Consortium (DPAC, \url{https://www.cosmos.esa.int/web/gaia/dpac/consortium}). Funding for the DPAC has been provided by national institutions, in particular the institutions participating in the {\it Gaia} Multilateral Agreement. 
The work of AVB, DAZ and YAS (optical data reduction) was supported by the baseline project FFUG-2024-0002 of the Ioffe Institute.
The analysis of the X-ray data by AVK was supported by the Russian Science Foundation project 22-12-00048-P. AK acknowledges the DGAPA-PAPIIT grant IA105024. 
SVZ acknowledges the DGAPA-PAPIIT grant IN119323. 
DAZ thanks Pirinem School of Theoretical Physics for hospitality. 
\end{acknowledgements}

\bibliographystyle{aa}
\bibliography{main}

\end{document}